\documentclass[12pt]{iopart}

\usepackage[english]{babel}

\usepackage{iopams,bm,graphicx,verbatim,mathrsfs,units,color}

\usepackage[colorlinks=true,citecolor=blue]{hyperref}

\newcommand{\ket}[1]{\left| #1 \right \rangle}

\newcommand{\figref}[1]{Fig.~\ref{#1}}

\begin{document}
\title{Single-electron entanglement and nonlocality}
\date{\today}

\author{David Dasenbrook$^1$, Joseph Bowles$^1$, Jonatan Bohr Brask$^1$, Patrick P. Hofer$^1$, Christian Flindt$^2$, and Nicolas Brunner$^1$}
\address{$^1$ D\'epartement de Physique Th\'eorique, Universit\'e de Gen\`eve, 1211
  Gen\`eve, Switzerland}
\address{$^2$ Department of Applied Physics, Aalto University,
  00076 Aalto, Finland}

\begin{abstract}
Motivated by recent progress in electron quantum optics, we revisit the question of single-electron entanglement, specifically whether the state of a single electron in a superposition of two separate spatial modes should be considered entangled. We first discuss a gedanken experiment with single-electron sources and detectors, and demonstrate deterministic (i.~e.~without post-selection) Bell inequality violation. This implies that the single-electron state is indeed entangled and, furthermore, nonlocal. We then present an experimental scheme where single-electron entanglement can be observed via measurements of the average currents and zero-frequency current cross-correlators in an electronic Hanbury Brown-Twiss interferometer driven by Lorentzian voltage pulses. We show that single-electron entanglement is detectable under realistic operating conditions. Our work settles the question of single-electron entanglement and opens promising perspectives for future experiments.
\end{abstract}



\maketitle

\section{Introduction}

The field of electron quantum optics has witnessed strong experimental advances over a short period of time \cite{bocquillon14}.  Electronic analogues of the Mach-Zehnder \cite{ji03}, Hanbury Brown-Twiss \cite{neder07} and Hong-Ou-Mandel interferometers \cite{bocquillon13} can now be implemented with edge channels of the integer quantum hall effect functioning as wave guides for electrons. At the same time, the recent realization of coherent single-electron emitters is opening up avenues for the controlled manipulation of few-particle electronic states \cite{feve07,fletcher13,dubois13nature,jullien14}. In parallel to these developments, a number of theoretical proposals have been put forward to entangle electrons, e.~g.~in edge channels \cite{lesovik01,oliver02,beenakker03,samuelsson04,scarani04}, using either the electron spin or the orbital degrees of freedom. The entanglement is detected by violating a Bell inequality \cite{bell64,brunner14} formulated in terms of zero-frequency current cross-correlations \cite{kawabata01,samuelsson03,samuelsson09}. While early proposals focus on electron sources driven by static voltages, more recent works investigate the on-demand generation of entangled states using dynamic single-electron emitters \cite{splettstoesser09,sherkunov12,hofer13,vyshnevyy13,dasenbrook15}.

\begin{figure}[h!]
  \centering
  \includegraphics[width=0.45\columnwidth]{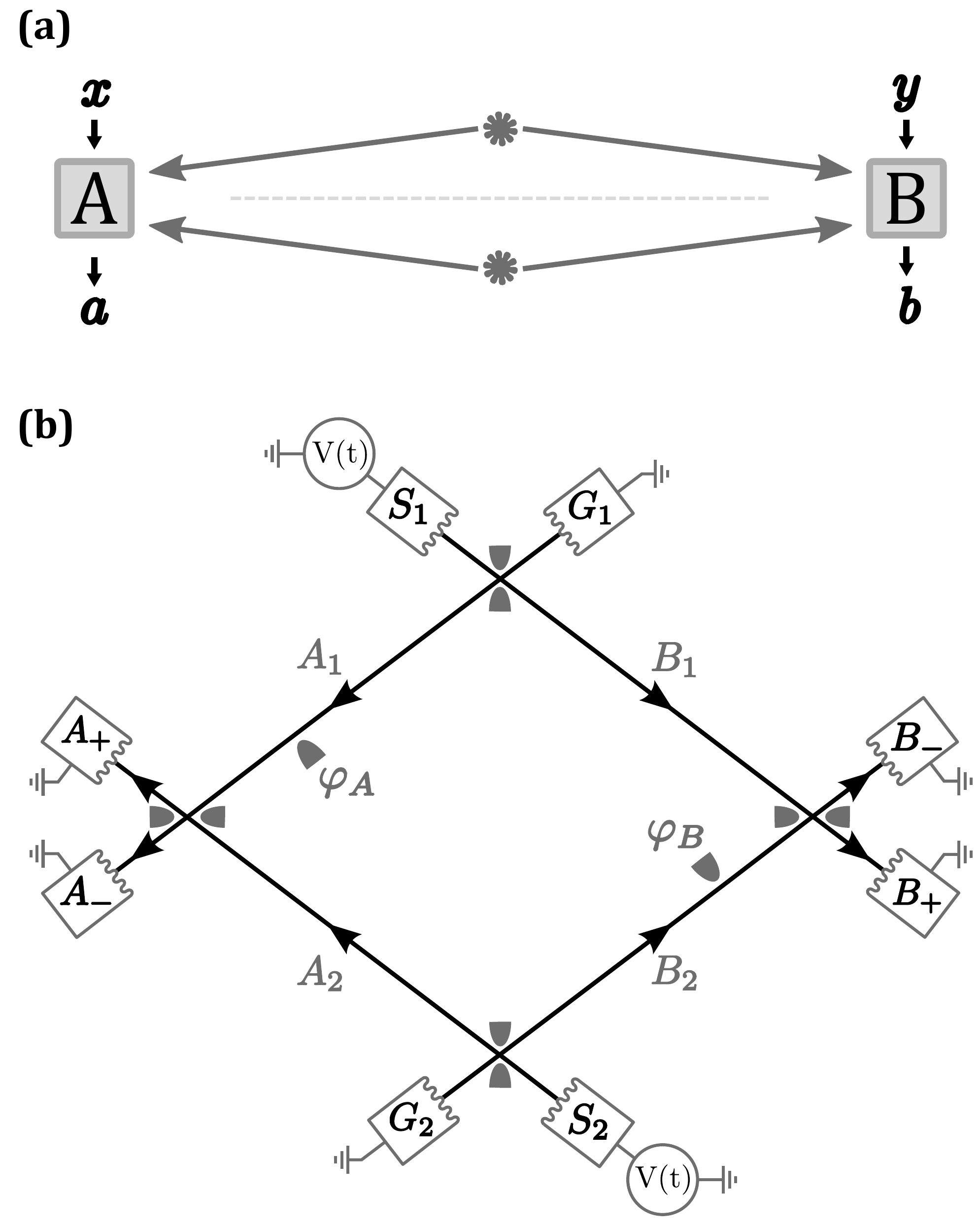}
  \caption{Schematic setup. \textbf{(a)} Two independent single-electron sources emit delocalized electrons towards the locations $A$ and $B$. A Bell test is performed using local operations and measurements at $A$ and $B$. If the resulting data $p(ab|xy)$ violates a Bell inequality, $A$ and $B$ necessarily share entanglement. Hence, the sources must emit entangled states. \textbf{(b)} Electronic Hanbury Brown-Twiss interferometer realizing the idea in (a) for an experimental demonstration of single-electron entanglement. Single-electron excitations are generated at the source contacts $S_1$ and $S_2$ and travel to the outputs $A_\pm$ and $B_\pm$. The contacts $G_1$ and $G_2$ are grounded.}
  \label{fig:idea_setup}
\end{figure}

For spin or orbital entanglement, several particles are involved and the particles are entangled in the spin or the orbital degrees of freedom, respectively. A conceptually different notion of entanglement is provided by entangled states of different occupation numbers. In this case, the entanglement is between different modes, and the relevant degree of freedom is the particle number in each mode. It is a question that has been much debated whether a state of a single particle in a superposition of two spatially separate modes should be considered entangled \cite{tan91,hardy94,ghz95,vaidman95,hardy95reply,wiseman03,bartlett07}. For photons (and other bosons) it is by now well established that the answer is yes, and that the entanglement is in fact useful in quantum communication applications \cite{vanenk05,sangouard12}. For electrons (and other fermions), the situation is different because of charge and parity superselection rules, and the question still causes controversy \cite{lebedev04,wiseman04,samuelsson05njp,giovanetti07,sherkunov09}.

Here, we revisit this question motivated by the recent development of dynamic single-particle sources in electron quantum optics. We demonstrate rigorously that the answer for electrons is affirmative based on the situation sketched in \figref{fig:idea_setup}(a): Two independent sources each produce a single electron which is delocalised with one part transmitted to location $A$ and the other to $B$. Using only local operations (LOs) and measurements at each location, a Bell inequality between $A$ and $B$ is violated deterministically, i.~e.~without post-selection. This necessarily implies that there is entanglement between $A$ and $B$. Since the sources are independent this in turn implies that the state emitted by a single source is entangled between regions $A$ and $B$. Specifically, we show that such a situation can be realized in an electronic Hanbury Brown-Twiss interferometer driven by Lorentzian voltage pulses as illustrated in \figref{fig:idea_setup}(b). Notably, the single-electron entanglement can be observed from current cross-correlation measurements at the outputs of the interferometer.

\section{Single-particle entanglement}

We start with a brief introduction to single-particle entanglement. A single particle in a superposition of two different locations can be described by the state
\begin{equation}
  \label{eq:splitsingleparticle}
  \ket{\Psi} = \frac{1}{\sqrt{2}} \left(\ket{0}_A\ket{1}_B + \ket{1}_A\ket{0}_B \right) ,
\end{equation}
where the numbers in the kets indicate the particle numbers in the spatially separated modes. The basic question is whether such a state is entangled. One can ask the question both for bosons and for fermions, in particular for photons and electrons. To answer affirmatively, the entanglement must be experimentally detectable.

Entanglement should be verified directly from measurements on each spatial mode in \Eref{eq:splitsingleparticle}, e.~g.~by testing the observations against a Bell inequality \cite{bell64,brunner14}. If arbitrary measurements were possible, \Eref{eq:splitsingleparticle} should indeed be considered
entangled since it for example violates the Clauser-Horne-Shimony-Holt (CHSH) Bell inequality \cite{clauser69}. However, the possible measurements may be limited because the state~\eref{eq:splitsingleparticle} is a single-particle state. Violating the CHSH inequality requires measurements which are not diagonal in the occupation number
basis, i.~e.~they should contain projections onto superpositions of states with different particle numbers  such as $(\ket{0} + \ket{1})/\sqrt{2}$. One may therefore expect a fundamental difference between photons and electrons because global charge conservation and parity superselection \cite{friis15,amosov15} forbids such superpositions for electrons \cite{schuch04,bartlett07}.

For photons it is by now established that the state given in~\Eref{eq:splitsingleparticle} is entangled and in fact useful for applications in quantum communication \cite{sangouard11,sangouard12}. Experimental demonstrations of single-photon entanglement have been reported using homodyne \cite{babichev04,fuwa15} and weak displacement measurements \cite{hessmo04,monteiro15}. These measurements require the use of coherent states of light (laser light), which introduces additional particles. These particles provide a reference frame between the observers \cite{bartlett07,brask13}. Alternatively, single-photon entanglement can be converted into entanglement between two atoms \cite{vanenk05}. In \Eref{eq:splitsingleparticle}, the numbers 0, 1 then represent internal atomic states and entanglement can be verified
straightforwardly. Importantly, since the conversion process involves only LOs, one concludes that the original single-photon state given in \Eref{eq:splitsingleparticle} must have been entangled. These procedures, however, cannot be straightforwardly applied to fermions (for example, there is no equivalent of coherent states for fermions). Hence, a more careful analysis is necessary as we show in the following.

\section{Single-electron entanglement and nonlocality}
\label{sec:revealing}

We consider the experiment pictured in \figref{fig:idea_setup}(b) and now argue that single-electron entanglement is observable. To keep the analysis simple, we work at zero temperature and assume that the sources create single electronic excitations above the Fermi sea which can be
detected one by one. These assumptions do not contradict any fundamental principle such as charge conservation. We consider the possibility of an experimental implementation with current technology later on.

Single electrons are excited above the Fermi sea at the sources $S_1$ and $S_2$, and are coherently split and interferred on electronic beamsplitters -- quantum point contacts (QPCs) tuned to half transmission. Tunable phases $\varphi_A$ and $\varphi_B$ can be applied in one arm on either side of the interferometer. The phases can be tuned using side gates or by changing the magnetic flux $\Phi$ through the device. In the latter case, we have $2 \pi \Phi/\Phi_0 = \varphi_A + \varphi_B$, where $\Phi_0=h/e$ is the magnetic flux quantum.

Labelling the modes as indicated in the figure, in second quantized notation the top beam splitter implements the transformation $a_{S_1}^\dagger \rightarrow (a_{A_1}^\dagger + a_{B_1}^\dagger)/\sqrt{2}$, $a_{G_1}^\dagger \rightarrow (a_{A_1}^\dagger - a_{B_1}^\dagger)/\sqrt{2}$ and similarly for the others. Here, we have introduced the fermionic creation and annihilation operators $a_\alpha^\dagger$ and $a_\alpha$ for electrons above the Fermi sea in mode $\alpha$. Considering just the top source ($S_1$), the state created after the beam splitter is thus
\begin{equation}
\label{eq:splitelectron}
\frac{1}{\sqrt{2}} (a_{A_1}^\dagger + a_{B_1}^\dagger) \ket{0},
\end{equation}
where the state $\ket{0}$ represents the undisturbed Fermi sea. This is the electronic version of \Eref{eq:splitsingleparticle}, and we use the interferometer to demonstrate that the state indeed is entangled between the regions $A$ and $B$.

The joint initial state of the two sources is $a_{S_1}^\dagger a_{S_2}^\dagger\ket{0}$, and the state evolution up to the output of the interferometer is then
\begin{eqnarray}
&a_{S_1}^\dagger a_{S_2}^\dagger \ket{0}  \rightarrow \frac{1}{2} (a_{A_1}^\dagger e^{i\varphi_A} +
a_{B_1}^\dagger)(a_{A_2}^\dagger + a_{B_2}^\dagger e^{i\varphi_B}) \ket{0} \nonumber \\
&\qquad \rightarrow \frac{1}{4} \bigg[ a_{A_+}^\dagger a_{B_+}^\dagger (e^{i\varphi}-1) +
a_{A_+}^\dagger a_{B_-}^\dagger (e^{i\varphi}+1) \nonumber \\
&\qquad \qquad + a_{A_-}^\dagger a_{B_+}^\dagger (e^{i\varphi}+1) + a_{A_-}^\dagger a_{B_-}^\dagger
(e^{i\varphi}-1) \nonumber \\
&\qquad \qquad - 2 e^{i\varphi_A} a_{A_+}^\dagger a_{A_-}^\dagger + 2 e^{i\varphi_B} a_{B_+}^\dagger
a_{B_-}^\dagger \bigg] \ket{0},
\label{eq:psiout}
\end{eqnarray}
where $\varphi = \varphi_A + \varphi_B$ and we have used the fermionic anti-commutation relations $\{a_i^\dagger, a_j\} = \delta_{ij}$ and $\{a_i^\dagger,a_j^\dagger\} = \{a_i,a_j\} = 0$. We omit terms where two electrons go to the same output since these are ruled out by the Pauli exclusion principle.\footnote{Such terms vanish due to the fermionic anti-commutation relations, e.~g.~$2(a_{A_1}^\dagger)^2 = \{a_{A_1}^\dagger,a_{A_1}^\dagger\} = 0$.}

Assuming that single-electron detection is possible, the state given in~\Eref{eq:psiout} can be seen to violate the CHSH inequality using the following strategy: The phases $\varphi_A^x$ and $\varphi_B^y$ are determined by the inputs $x,y=0,1$, and the binary outputs $a,b=\pm 1$ are determined by outputting $\pm 1$ when one click is observed in detector $A_\pm$ (similarly for $B$). In cases where both or none of the detectors click, the outputs are defined to be $+1$ and $-1$ respectively. We denote the probability for outputs $a$, $b$ given inputs $x$, $y$ by $P(ab|xy)$. The correlator defined as
\begin{equation}
E_{xy} = \sum_{a,b} a b P(ab|xy)
\end{equation}
is then given by
\begin{equation}
 \label{eq:Exycorrelator}
E_{xy}  = -\frac{1+\cos(\varphi_A^x + \varphi_B^y)}{2} .
\end{equation}
If the experiment can be explained by a local hidden variable model, then the CHSH inequality holds \cite{clauser69}
\begin{equation}
S = | E_{00} + E_{01} + E_{10} - E_{11} | \leq 2 .
\end{equation}
Now, with the choice $\varphi_A^0=0$, $\varphi_A^1=\pi/2$, $\varphi_B^0=-3\pi/4$, and
$\varphi_B^1=3\pi/4$, we find
\begin{equation}
S = 1+\sqrt{2} > 2 .
\end{equation}
Thus, the CHSH inequality is clearly violated. Since the state given in \Eref{eq:psiout} violates a Bell inequality between $A$ and $B$, it must necessarily be entangled. Note that this Bell inequality violation is not subjected to the detection loophole \cite{brunner14}, as our scheme does not involve any post-selection. Furthermore, the state given in \Eref{eq:psiout} was created by LOs on two copies of the state given in \Eref{eq:splitelectron} coming from two independent sources. Since any product of separable states is separable, it follows that the state given in \Eref{eq:splitelectron} must itself be entangled. We thus conclude that the state of a single electron split between two modes is entangled.

It should be pointed out that the setup in \figref{fig:idea_setup}(b) is similar to the Hanbury Brown-Twiss interferometer for electrons, as theoretically proposed \cite{samuelsson04} and experimentally realized \cite{neder07} using edge states of a two-dimensional electron gas in the
integer quantum hall regime. However, in these works maximal CHSH inequality violation ($S=2 \sqrt{2}$) is achieved by post-selection on the subspace of one electron on each side of the interferometer (effectively post-selecting a maximally entangled state), which is interpreted as two-electron orbital entanglement. Here, by contrast, our scheme involves no post-selection and we do not achieve maximal CHSH violation, but in turn we can demonstrate single-electron entanglement.

It should also be noted that the possibility of using two copies of a single electron entangled state in order to distill one entangled two-electron state has been discussed in
Refs.~\cite{wiseman03,vaccaro03}. There, the idea is that each observer performs a non-demolition measurement of the local electron number and then post-selects on the cases where a single electron is detected on each side. Alternatively, the distilled entanglement can be transferred to a pair of additional target particles \cite{ashhab07}, in which case however single-electron nonlocality cannot be unambiguously concluded. Again, as argued above, our setup involves no post-selection and is thus conceptually different. Moreover, the setup does not require non-demolition measurements.

The scheme described so far is a thought experiment, demonstrating that single-electron entanglement in theory is observable. In principle, nothing prevents its realization. Single-electron sources  \cite{feve07,dubois13nature,jullien14} and electronic beam splitters have been experimentally realized and the first steps towards single-electron detectors \cite{thalineau14,fletcher13} have recently been taken. Still, realizing our thought experiment is at present challenging, mainly because of the requirement to detect single electrons. To relax this constraint, we discuss in the next section an experiment which only relies on measurements of the average current and the zero-frequency current-correlators. These are standard measurements which would also demonstrate single-electron entanglement, albeit under slightly stronger assumptions about the experimental implementation.

\section{Observing single-electron entanglement}
\label{sec:currentnoise}

We consider again the setup in \figref{fig:idea_setup}(b), but now discuss a detection scheme which is feasible using existing technology. Specifically, we consider measurements of zero-frequency currents and current correlators as an alternative to single-electron detection. We give a detailed description of the single-electron sources and the interferometer based on Floquet scattering theory \cite{pedersen98,moskalets02,moskaletsbook,dubois13}. This allows us to investigate realistic operating conditions such as finite electronic temperatures and dephasing. As we will see, it is possible to demonstrate single-electron entanglement under one additional assumption, namely that the measurement of the mean current and the zero-frequency current correlators amounts to taking ensemble averages over the state in each period of the driving. This is a reasonable assumption if the period of the driving is so long that only one electron from each source is traversing the interferometer at any given time.

For the single-electron sources, we consider the application of Lorentzian-shaped voltage pulses to the contacts \cite{levitov96,ivanov97,lebedev05,keeling06,dubois13nature,jullien14}. A driven mesoscopic capacitor \cite{feve07} can be used instead. Electrons leaving a contact pick up a time-dependent phase
\begin{equation}
  \label{eq:timedepphase}
  \varphi(t) = - \frac{e}{\hbar} \int_{-\infty}^t V(t') \mathrm{d} t',
\end{equation}
where the voltage applied to the contact has the form
\begin{equation}
  \label{eq:lorentzianvoltage}
  eV(t) = \sum_{j=-\infty}^{\infty} \frac{2\hbar \Gamma}{\left(t-n\mathcal{T} \right)^2 + \Gamma^2}.
\end{equation}
At zero temperature, this results in the excitation of exactly one electron out of the Fermi sea (and one hole going into the contact) without any additional electron-hole pairs. This quasiparticle is called a leviton \cite{dubois13nature,jullien14}. In \Eref{eq:lorentzianvoltage}, the temporal width of the pulse is denoted as $\Gamma$ and
$\mathcal{T}$ is the period of the driving.

Floquet scattering theory provides us with a convenient theoretical framework to describe the periodically driven interferometer \cite{pedersen98,moskalets02,moskaletsbook,dubois13}. By Fourier transforming \Eref{eq:timedepphase}, we obtain the Floquet scattering matrix of the driven contacts as
\begin{equation}
  \label{eq:floquetsmatrixcontact}
  \mathcal{S}_l(n) = \cases{
    -2 e^{-n\Omega \Gamma} \sinh(\Omega \Gamma) & $n > 0$ \\
    e^{-\Omega \Gamma} & $n=0$ \\
    0 & $n < 0$}.
\end{equation}
These are the amplitudes for an electron at energy $E$ to leave the contact at energy $E_n = E + n \hbar \Omega$, having absorbed ($n>0$) or emitted ($n<0$) $|n|$ energy quanta of size $\hbar\Omega$, where $\Omega = 2 \pi / \mathcal{T}$ is the frequency of the driving.

The scattering matrix of the interferometer can be found as follows. Since there are eight terminals in total (four inputs and four outputs), the scattering matrix of the interferometer is an $8 \times 8$ matrix. However, due to the chirality of the edge states, electrons leaving an input contact can only travel to an output. This allows us to work with an effective $4 \times 4$ scattering matrix connecting every possible input to every possible output. Including the phases $\varphi_A$
and $\varphi_B$, that the particles pick up when travelling from input 1 to location $A$ or from input 2
to $B$, the scattering matrix reads
\begin{equation}
  \label{eq:stationarySmatrix}
  \mathcal{S} = \left( \begin{array}{cccc}
    r_1 r_A e^{i \varphi_A} & r_1 t_A e^{i \varphi_A} & t_1 t_B & t_1 r_B \\
    t_1 r_A e^{i \varphi_A} & t_1 t_A e^{i \varphi_A} & -r_1 t_B & -r_1 r_B \\
    t_2 t_A & - t_2 r_A & -r_2 r_B e^{i \varphi_B} & r_2 t_B e^{i \varphi_B} \\
    -r_2 t_A & r_2 r_A & -t_2 r_B e^{i \varphi_B} & t_2 t_B e^{i \varphi_B}
  \end{array} \right).\nonumber
\end{equation}
Here, $t_{1(2)}$ refers to the transmission amplitude of the QPCs after source $1(2)$ and $t_{A(B)}$ is the amplitude for the QPC located at $A(B)$. The $r$'s are the corresponding reflection amplitudes. The rows number the possible inputs $S_1$, $G_1$, $S_2$ and $G_2$ (in this order) and the columns the possible
outputs $A+$, $A-$, $B+$ and $B-$. We have chosen all amplitudes to be real and inserted factors of $-1$ for half of the reflection amplitudes to ensure the unitarity of the scattering matrix. Below, we consider only half-transparent beam splitters and thus set all amplitudes to $1/\sqrt{2}$.

To obtain the combined Floquet scattering matrix of the interferometer and the single-electron
sources, we multiply every matrix element of the stationary $\mathcal{S}$-matrix corresponding to a
voltage-biased input (i.~e.~the first and third rows) by $\mathcal{S}_l(n)$ and every element
corresponding to a grounded input (i.~e.~the second and fourth rows) by $\delta_{n0}$. In doing so,
we assume that the two electron sources are perfectly synchronized and all arms of the
interferometer have the same length. The resulting Floquet scattering matrix
$\mathcal{S}_F(E_n,E)\equiv\mathcal{S}_F(n)$ is the basis of all calculations below.

The current operator in output $\alpha$ is given by \cite{blanter00}
\begin{equation}
  \label{eq:currentoperator}
  I_\alpha = \frac{e}{h} \int_{-\infty}^\infty \left\{ c_\alpha^\dagger(E)
  c_\alpha(E) - b_\alpha^\dagger(E) b_\alpha(E) \right\} \mathrm{d} E,
\end{equation}
where the operators $c_\alpha(E)$ ($b_\alpha(E)$) annihilate an incoming (outgoing) electron in lead $\alpha$ at energy $E$. Outgoing electrons from the leads are distributed according to the Fermi-Dirac distribution function
\begin{equation}
  \label{eq:fermidirac}
  \langle b^\dagger_\alpha(E) b_\beta(E') \rangle = \delta_{\alpha \beta} \delta(E-E')
  \frac{1}{e^{E/(k_B T)}+1},
\end{equation}
where $T$ is the electronic temperature and we have set the Fermi level in all reservoirs to $E_F = 0$. The scattered electrons are not in thermal equilibrium. We find their distribution by relating the incoming electrons to the outgoing ones via the Floquet scattering matrix as \cite{moskaletsbook}
\begin{equation}
  \label{eq:incomingoutgoing}
  c_\alpha(E) = \sum_{n=-\infty}^\infty \sum_\beta [\mathcal{S}_F(E,E_n)]_{\alpha \beta}
  b_\beta(E_n).
\end{equation}

\subsection{Zero temperature}
At zero temperature, the average currents and the zero-frequency current correlators can be calculated analytically using \Eref{eq:currentoperator} and \eref{eq:incomingoutgoing}. For example, the average current at output $A+$ reads
\begin{equation}
  \label{eq:currentA+}
  \langle I_{A+} \rangle = \frac{e}{\mathcal{T}} \left( T_2 T_A + T_1 R_A \right),
\end{equation}
where $T_i = |t_i|^2$ and $R_i = |r_i|^2$ ($i = 1,2,A,B$). The zero-frequency current cross-correlator is defined as
\begin{equation}
  \label{eq:zerofreqnoise}
  P_{\alpha \beta} = \left \langle I_\alpha I_\beta \right \rangle - \left \langle I_\alpha \right
  \rangle \left \langle I_\beta \right \rangle.
\end{equation}
For the cross-correlator between the $A+$ and $B+$ outputs we obtain
\begin{equation}
  \label{eq:noiseA+B+}
  P_{A+B+} = - \frac{e^2}{\mathcal{T}} \left| t_2 t_A r_2 t_B e^{i \varphi_B} + t_1 r_A r_1 r_B
  e^{-i\varphi_A} \right|^2.
\end{equation}
Note that the average currents are insensitive to the phases $\varphi_A$ and $\varphi_B$, whereas the current cross-correlators depend on their sum $\varphi_A+\varphi_B$. This is known as the two-particle Aharonov-Bohm effect \cite{samuelsson04}.

We now formulate the CHSH inequality \cite{clauser69} for our system. The leviton annihilation operator is \cite{keeling06}
\begin{equation}
  \label{eq:levitonoperator}
  a_\alpha = \sqrt{2 \Gamma} \sum_{E>0} e^{-\Gamma E/\hbar} b_\alpha(E).
\end{equation}
At zero temperature, we can express the operator of the number of levitons emitted from lead $\alpha$ per period in terms of the current operator as
\begin{equation}
  \label{eq:levitonnumber}
  a_\alpha^\dagger a_\alpha = \frac{\mathcal{T}}{e} I_\alpha.
\end{equation}
This allows us to relate the current operator for a given detector at $A$ or $B$ to an operator on the modes on side $A$ or $B$ before the final beam splitter and the phase shift, cf.~\figref{fig:idea_setup}(b). Taking for instance the detector $A_+$ and transforming \Eref{eq:levitonnumber} through the beam splitter and the phase shift, we get
\begin{eqnarray}
\label{eq:transformednumberop}
a_{A_+}^\dagger a_{A_+} \rightarrow & \frac{1}{2} ( e^{-i\varphi_A} a_{A_1}^\dagger + a_{A_2}^\dagger )( e^{i\varphi_A} a_{A_1} + a_{A_2} ) \nonumber\\
 =& \frac{1}{2} ( a_{A_1}^\dagger a_{A_1} + a_{A_2}^\dagger a_{A_2}) + \frac{1}{2} (e^{-i\varphi_A} a_{A_1}^\dagger a_{A_2} + e^{i\varphi_A} a_{A_2}^\dagger a_{A_1} ) .
\end{eqnarray}

To gain an intuitive understanding of this operator, we consider its restriction to the
single-electron subspace, i.~e.~the case where there is exactly one electron on side $A$ of the
interferometer. In this case, the first term in \eref{eq:transformednumberop} is just $1/2$. The
Hilbert space is two-dimensional and the states $a_{A_1}^\dagger\ket{0}$, $a_{A_2}^\dagger\ket{0}$
form a basis. In this basis, the second term in \eref{eq:transformednumberop} is $(\cos(\varphi_A)
\sigma_x + \sin(\varphi_A) \sigma_y)/2$, with $\sigma_x$, $\sigma_y$, $\sigma_z$ being the usual
Pauli matrices. Thus, in the single-electron subspace we have
\begin{equation}
  I_{A_+} = \frac{e}{2\mathcal{T}} \left( 1 + \sigma^A_{\varphi_A} \right) ,
\end{equation}
where $\sigma^A_{\varphi_A} = \cos(\varphi_A) \sigma^A_x + \sin(\varphi_A) \sigma^A_y$ is the rotated Pauli matrix in the $x$-$y$ plane, acting on side $A$. From this we see
that, in the single-electron subspace, measuring $I_{A_+}$ is equivalent to measuring $\sigma^A_{\varphi_A}$. Similar expressions can be obtained for the currents at the other detectors, and thus, by measuring the currents at the four outputs, we can measure any combination of Pauli operators in the two-qubit subspace with a single electron on each side of the interferometer.

With this in mind, we define the observables
\begin{equation}
  \label{eq:Xoperators}
  X_A^{\varphi_A} = \frac{2\mathcal{T}}{e} I_{A_+}^{\varphi_A} - 1 , \qquad 
  X_B^{\varphi_B} = \frac{2\mathcal{T}}{e} I_{B_+}^{\varphi_B} - 1 ,
\end{equation}
where the current for a given phase setting $\varphi$ is denoted as $I_\alpha^\varphi$. In the subspace with one electron on each side of the interferometer, these correspond to measuring (rotated) Pauli operators. Events where two or no electrons arrive on the same side will give contributions of $+1$ or $-1$ respectively, cf.~\Eref{eq:levitonnumber}, independent of the phase settings, analogously to the output strategy in the previous section. At zero temperature the
correlator becomes
\begin{equation}
\label{eq:currentcorr}
\langle X_A^{\varphi_A} X_B^{\varphi_B} \rangle = - \frac{1+\cos(\varphi_A+\varphi_B)}{2},
\end{equation}
showing that the joint statistics is the same as in Sec.~\ref{sec:revealing}, where single-electron detection was assumed. Here, however, we interpret the current expectation values entering in the correlator, such as $\langle I_{A_+}^{\varphi_A} \rangle$, as the result of time-integrated measurements. We thus assume that a measurement of the time-integrated current and the zero-frequency current correlators amounts to taking ensemble averages over the state in each period of the driving. The statistics obtained from the time-integrated current measurement is then the same as what one would obtain by averaging over several periods of the driving with single-electron detection. Under this assumption, we can again consider the CHSH inequality
\begin{equation}
\label{eq:chsh}
S = \Big| \Big \langle X_A^{\varphi_A^0} X_B^{\varphi_B^0} + X_A^{\varphi_A^0} X_B^{\varphi_B^1} + X_A^{\varphi_A^1} X_B^{\varphi_B^0} - X_A^{\varphi_A^1} X_B^{\varphi_B^1} \Big \rangle \Big| \leq 2,
\end{equation}
It is easy to see that the choice $\varphi_A^0 = 0$, $\varphi_A^1 = \pi/2$, $\varphi_B^0 = -3\pi/4$, $\varphi_B^1 = 3\pi/4$ leads to a violation, giving
\begin{equation}
\label{violation}
S = 1+\sqrt{2} > 2.
\end{equation}
This finally shows us that this scheme makes it possible to observe single-electron entanglement using zero-frequency measurements only.

We note that our results for the current and the zero-frequency noise do not depend on the pulse width $\Gamma$. As such, our measurement strategy based on \Eref{eq:Xoperators} would also work with constant voltages as realized in the experiment by Neder \emph{et al.}~\cite{neder07}, and the CHSH violation of \Eref{violation} would be obtained. However, to unambiguously demonstrate single-electron entanglement, in line with the thought experiment described in Sec.~\ref{sec:revealing}, it is important that only one electron from each source is traversing the interferometer at any given time. We therefore need to work with a long period and well-separated pulses, as opposed to constant voltages.

It is instructive to compare our proposal to the previous work of Samuelsson \emph{et al.}~\cite{samuelsson04}. Although the two setups are similar, the detection scheme discussed here is different. This significantly changes the interpretation of the observations. The measurement scheme
suggested by Samuelsson \emph{et al.}~is formulated in terms of coincidence rates \cite{samuelsson04,samuelsson09ps}. The corresponding observable is then sensitive only to the part of the state with a single electron on each side of the interferometer. Thus, the measurement effectively corresponds to performing post-selection, discarding the part of the state where two electrons are on the same side. In this case, the CHSH inequality is maximally violated ($S=2\sqrt{2}$), as the post-selected state is a maximally entangled two-qubit state. The Bell inequality is then violated because of the two-electron orbital entanglement \cite{samuelsson04}. By contrast, our measurement strategy is sensitive to the entire state (including terms with two
electrons on the same side) and does not imply any effective post-selection. For this reason we reach a lower CHSH violation, $S=1+\sqrt{2}$. However, we observe in turn single-electron entanglement.

\subsection{Finite temperatures and dephasing}

At finite temperatures, additional excitations in terms of electron-hole pairs are expected. Consequently, \Eref{eq:levitonnumber} does not hold any longer. The operators in \Eref{eq:Xoperators} are thus not strictly bounded between -1 and +1, although values outside this range should be rare at low temperatures. Since the CHSH parameter $S$ is a monotonically decreasing function of temperature, a violation of the CHSH inequality at finite temperatures indicates that the corresponding zero temperature state is unambiguously entangled. We will thus continue to use \Eref{eq:chsh} to detect single-particle entanglement.

\begin{figure}
  \centering
  \includegraphics[width=.8\columnwidth]{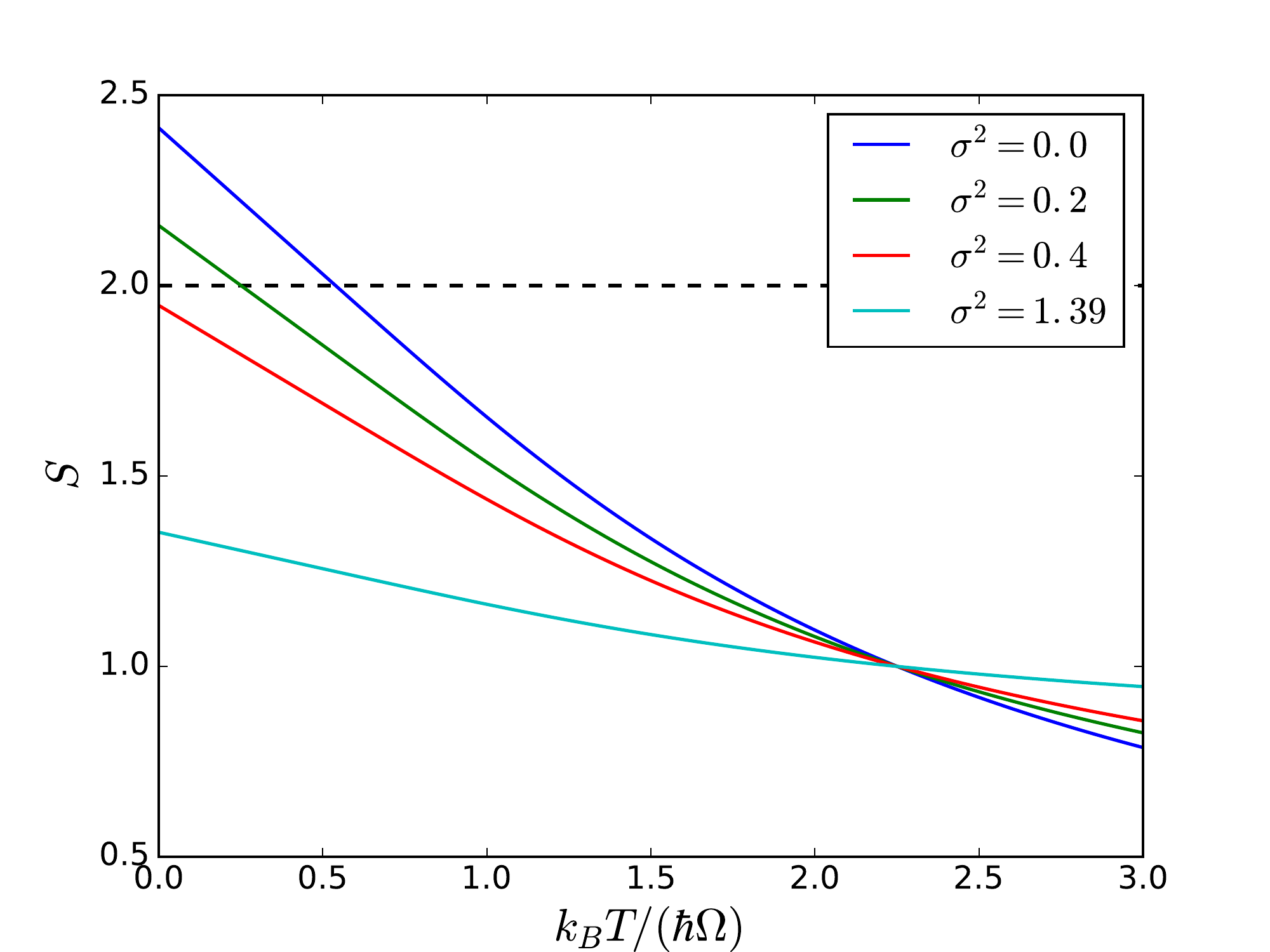}
  \caption{Maximal value of the CHSH parameter as a function of temperature. The Bell angles are $\varphi_A=0$, $\varphi_A'=\pi/2$, $\varphi_B=-\pi/4$ and $\varphi_B'=5\pi/4$. The dephasing parameter $\sigma^2$ is the variance of the distribution of the sum of the phases $\varphi_A+\varphi_B$. The dashed line indicates the CHSH bound.}
  \label{fig:chshtemperature}
\end{figure}

At finite temperatures, the average current and the zero-frequency current correlators can be calculated numerically. Figure \ref{fig:chshtemperature} shows the maximal value of the CHSH parameter (using the same phase settings as above) as a function of the electronic temperature. In the absence of any additional dephasing mechanisms (blue curve), the CHSH inequality can be violated up to a temperature of $k_B T \approx 0.5 \hbar \Omega$. For a typical driving frequency of $5$ GHz \cite{dubois13nature,jullien14}, this corresponds to a temperature of about $120$ mK, which is well within experimental reach.

Due to interactions with the electrons in the underlying Fermi sea as well as with nearby conductors, the injected single-electron states may experience decoherence and dephasing. Here we do not give a microscopic model for theses interactions, but instead we introduce a phenomenological dephasing parameter $\sigma^2$ which denotes the variance of the total phase $\varphi_A+\varphi_B$ in a model that leads to Gaussian phase averaging. Previous experiments have shown that this is the dominant effect of the interaction of electronic interferometers with their environments \cite{ji03,roulleau07}. At zero temperature, the correlator in \Eref{eq:currentcorr} then becomes
\begin{equation}
  \langle X_A^{\varphi_A} X_B^{\varphi_B} \rangle = - \frac{1 +
   e^{-\sigma^2}\cos(\varphi_A+\varphi_B)}{2} ,
\end{equation}
making a Bell violation possible up to $\sigma^2 \lesssim 0.35$. At finite temperatures, an analogous expression can be found \cite{samuelsson09ps} and the dephasing has a similar qualitative effect. Figure~\ref{fig:chshtemperature} shows that for small values of the dephasing parameter, a CHSH violation is still possible at low enough temperatures, while for $\sigma^2 \gtrsim 0.35$, the entanglement cannot be detected any longer. We note that the visibility of the current correlators observed in the experiment by Neder \emph{et al.}~\cite{neder07} is too low to violate \Eref{eq:chsh} in this setup. It corresponds to a dephasing parameter of $\sigma^2 \approx 1.39$ (light blue line in
Fig.~\ref{fig:chshtemperature}). Nevertheless, by a careful design of the interferometer the dephasing may be further reduced, bringing the measurement described here within experimental reach.

\section{Conclusions}

We have revisited the question of single-electron entanglement. Specifically, we have demonstrated theoretically that the state of a single electron in a superposition of two separate spatial modes is entangled.  As we have shown, single-electron entanglement can in principle be observed in an electronic Hanbury Brown-Twiss interferometer based on single-electron sources, electronic beam splitters, and single-electron detectors. Unlike earlier proposals for generating entanglement in electronic conductors, our scheme does not rely on any post-selection procedures. Since single-electron detectors are still under development, we have devised an alternative experimental scheme based on existing technology using average current and cross-correlation measurements. With these developments, the experimental perspectives for observing single-electron entanglement seem promising.

\ack
DD and PPH gratefully acknowledge the hospitality of Aalto University and McGill University, respectively. CF is affiliated with Centre for Quantum Engineering at Aalto University. This work was supported by the Swiss National Science Foundation (grants 200020\_150082, PP00P2\_138917
and Starting Grant DIAQ) and the Academy of Finland.

\end{document}